\documentclass{article}
\usepackage{spconf,amsmath,graphicx}
\usepackage{float} 
\usepackage{subfigure} 
\usepackage{color,soul}
\usepackage{hyperref}
\hypersetup{
    colorlinks=true,
    linkcolor=blue,
    filecolor=magenta,      
    urlcolor=cyan,
    pdftitle={Overleaf Example},
    pdfpagemode=FullScreen,
    }

\title{CONTRAST-PLC: CONTRASTIVE LEARNING FOR PACKET LOSS CONCEALMENT}
\name{Huaying Xue, Xiulian Peng, Yan Lu}
\address{Microsoft Research Asia, Beijing, China}
\begin{document}
%
\maketitle
\begin{abstract}
Packet loss concealment (PLC) is challenging in concealing missing contents both plausibly and naturally when there are only limited available context to use. Recently deep-learning based PLC algorithms have demonstrated their superiority over traditional counterparts; but their concealment ability is still mostly limited to a maximum of 120ms loss. Even with strong GAN-based generative models, it is still very challenging to predict long burst losses that could happen within/in-between phonemes. In this paper, we propose to use contrastive learning to learn a loss-robust semantic representation for PLC. A hybrid neural PLC architecture combining the semantic prediction and GAN-based generative model is designed to verify its effectiveness. Results on the blind test set of Interspeech2022 PLC Challenge show its superiority over commonly used UNet-style framework and the one without contrastive learning, especially for the longer burst loss at \((120, 220]\)ms.
\end{abstract}

\section{Introduction}
\label{sec:intro}
In voice over IP (VOIP), speech packets are prone to suffer from many type of errors, e.g. packet losses, delay and network jitters. These errors, if not handled properly, will result in severe discontinuity which largely degrade the quality and intelligibility of the calls. So most modern real-time communication systems will integrate a packet loss concealment (PLC) module at the receiver side to recover the missing parts before playing out. Traditional PLC techniques simply repeat pitch periods or process a linear regression on the highly-correlated pitch periods for the voiced component in parametric domain \cite{stimberg2020waveneteq}. These signal processing based methods have yielded good audio quality for short packet losses of about 40ms but introduced mechanical noise and waveform attenuation for longer packet losses.

 \begin{figure}[tb]
\centering 
\includegraphics[width=0.5\textwidth]{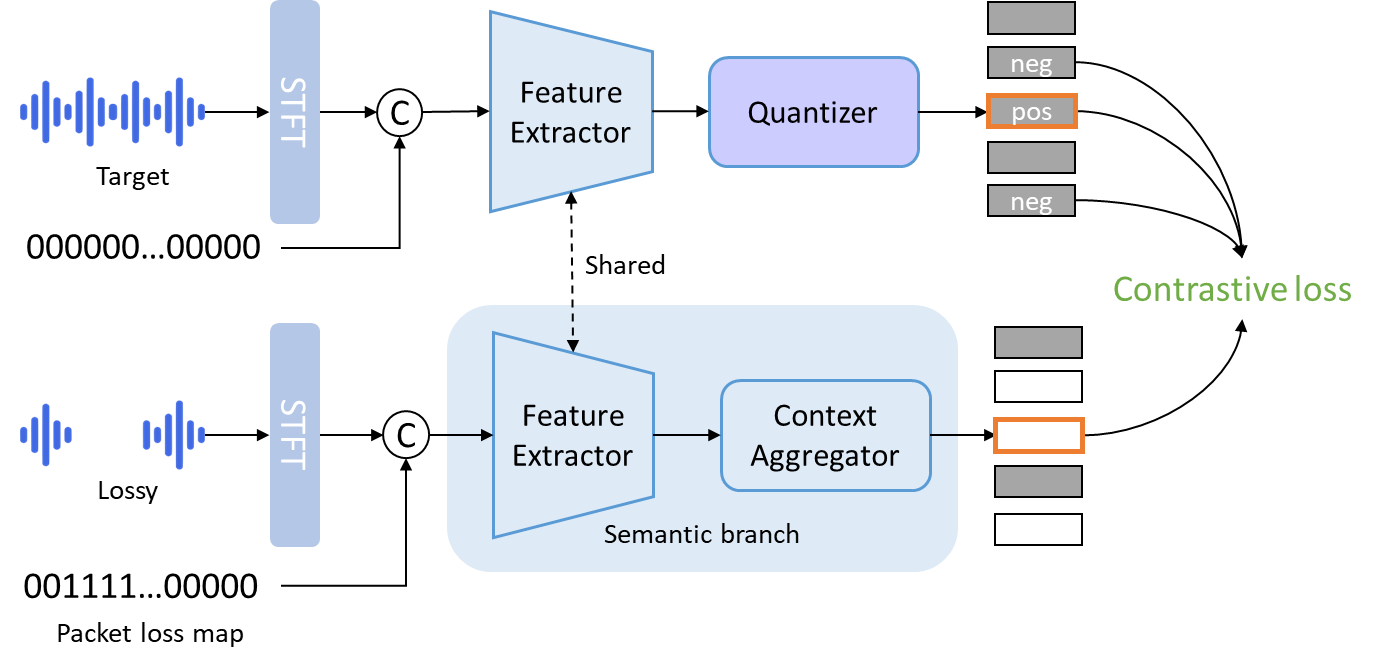} 
\caption{Contrastive learning for semantic representation learning for PLC.}
\vspace{-0.4cm}
\label{Fig_scheme_contrastive}
\end{figure}

 \begin{figure*}[htb] 
\centering 
\includegraphics[width=1.0\textwidth]{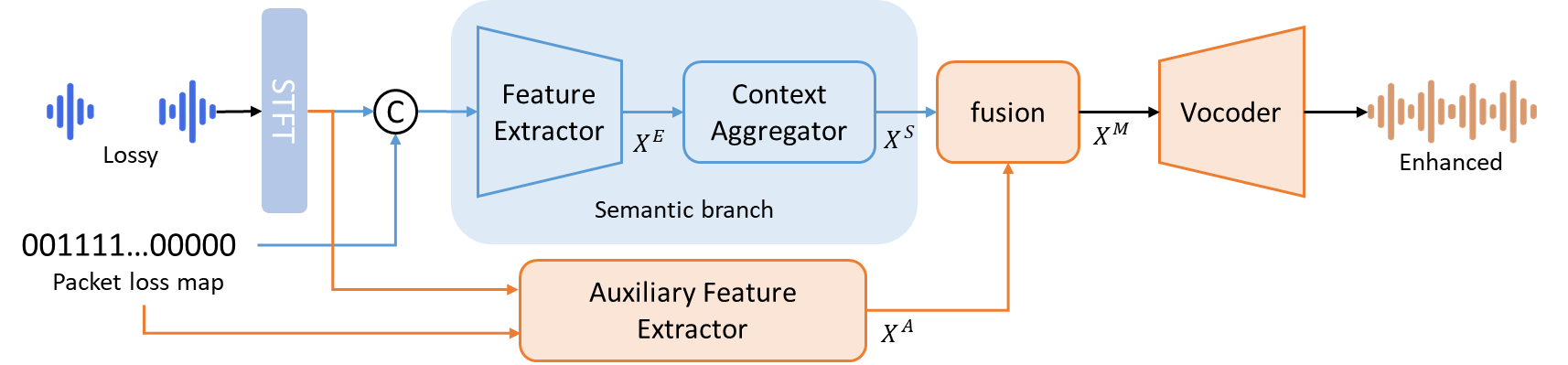} 
\vspace{-0.6cm}
\caption{The overall framework. Waveform is synthesized conditioned on the predicted semantic features and the auxiliary speech-related information. The semantic branch is pretrained by a contrastive task as shown in Fig.\ref{Fig_scheme_contrastive} and freezed in the downstream task, i.e. PLC, with adversarial training.}
\vspace{-0.6cm}
\label{Fig_scheme} 
\end{figure*}

Deep-learning based PLC algorithms have shown superior restoration ability for longer gaps, thanks to the emerging breakthroughs in speech synthesis. According to different technical stacks of speech synthesis, most PLC algorithms can be divided into auto-regressive networks\cite{stimberg2020waveneteq,lin2021time} and generative adversarial networks (GANs)\cite{binkowski2019high, pascual2021adversarial,wang2021temporal,liu2022plcnet,li2022end}. The autoregressive models use recurrent neural networks to process sample-wise regression which is time-consuming for training and inference. What's more, due to its generative property, it needs a special tunning of the sampling process to smooth the predicted samples with the real signals coming after the lost packets. In contrast, generative adversarial networks generate audio samples in parallel which has faster inference speed and is more friendly to highly-parallel computing devices. \cite{pascual2021adversarial} and \cite{liu2022plcnet} also show their superior quality over deep autoregressive models. For model structures, UNet-style architectures are widely used in the literature, either in time-domain \cite{wang2021temporal,liu2022plcnet,li2022end} or frequency-domain \cite{lim2018time}. This structure is elegant for light-weight models thanks to the skip connection aggregating information of different levels from encoder to decoder. Besides UNet, the prediction-and-synthesis based structure also shows promising results and its potential for more delicated control over different information. In this line, \cite{zhou2022neural} proposed to predict the mel-spectrum of the lost packet and transform it to waveform by a flow-based neural vocoder. Recent studies from Amazon \cite{valin2022real} predicted the acoustic features to be used by an LPCNet. These methods have been proven to be able to conceal reliable and natural contents for packet losses at the maximum of 120ms but fail to predict contents for longer gaps, appearing within or in-between phonemes. 

Inspired by the achievements in self-supervised speech representation learning (SSL), in this paper we investigate on using contrastive learning to enhance the semantic-level prediction ability of PLC networks. Contrastive learning has been widely used in SSL to learn linguistic representations. The wav2vec 2.0 \cite{baevski2020wav2vec} employed a masking scheme on features of different time steps and exploited a contrastive task by identifying the true quantized contextualized representations from a set of distractors. The learned representations have demonstrated good capability in speech and phoneme recognition tasks. However, these works take original speech without any loss as input and the learned representations are verified in understanding tasks.

In this work, we investigate how to use contrastive learning to learn loss-robust linguistic features and enhance the semantic prediction ability for PLC task. A hybrid framework combining semantic prediction and GAN-based generative model is introduced to verify its effectiveness. We are targeting at improving the phoneme-level concealment on longer losses over 120ms but not introducing any unreliable or new phonemes/words. Experimental results show that the proposed scheme achieves both lower WER and better quality in concealing longer gaps from 120ms to 220ms than the UNet-style counterparts and the one without contrastive learning.


 \begin{table*}[h!t]
\centering
\vspace{-0.2cm}
\caption{Quality comparison on the blind test set with real traces. Except for MCD and WER, the higher the better.}
\begin{tabular}{ c|c|c|c|c|c|c}
\hline
Scheme&PESQ&MCD[dB]&PLCMOS&NISQA Discontinuity&NISQA TTS&	WER[\%]\\
\hline
baseline&	\textbf{2.879}&	\textbf{0.920}&	3.980&	3.635&	2.818&	6.440\\
proposed w.o contrastive&	2.679&	1.316&	4.057&	3.676&	3.035&	7.390\\
proposed&	2.730&1.284&	\textbf{4.096}&	\textbf{3.775}&	\textbf{3.250}&	\textbf{5.580}\\
\hline
\end{tabular}
\label{table:tabel1}
\end{table*}

\section{THE PROPOSED SCHEME}
\label{sec:scheme}
Fig.\ref{Fig_scheme} shows the overall framework of the proposed method. It takes the lossy audio and a packet loss map indicating which frame is lost as input. The input signal is transformed into time-frequency spectrum with a 20ms window and a 10ms hop length with power-law compressed normalization on the magnitude. The network is composed of two branches for feature extraction and a vocoder for synthesis. The semantic branch targets at extracting/predicting linguistic features, which is pre-trained by a contrastive task as shown in Fig.\ref{Fig_scheme_contrastive}. Besides linguistic features, we add an auxiliary branch to extract other low-level features which are necessary for a good synthesized audio. The output of these two branches are fused before feeding into the vocoder for synthesis. A modified causal HiFi-GAN\cite{kong2020hifi} is employed as the vocoder with adversarial training. The whole scheme takes a sequence-to-sequence manner without replacing the output with the input for non-lost frames, thanks to the good adaptation of the network. For long burst losses over 140ms, a smoothstep function based fade-in-fade-out operation is applied as post-processing to avoid artifacts and meaningless contents to be generated. The following subsections will describe them in detail.

\subsection{Network Structure}
\label{ssec:nework structure}
The semantic branch is composed of causal 2D convolutional blocks in feature extractor and groups of dilated causal temporal convolution modules (TCM) \cite{jiang2022end} for context aggregator. Let \(X^{E}\in{R}^{C_{E}\times T\times F}\) denote the output of the feature extractor. All the frequency information is folded into channels \(C_{S}=C_{E}\times F\) before feeding into the context aggregator, which finally produces an output \(X^{S}\in{R}^{C_{S}\times T}\).

The auxiliary branch extracts low-level features necessary to synthesize the audio in addition to the semantic branch. It consists of one causal 2D convolutional block followed by several modified group-wise temporal self-attention blocks (G-TSA) based on the work in \cite{xue2022towards}. As the lost frames typically have little or weak information in the extracted features, we employ a masking on the query, key and value of the G-TSA block where the lost frame is set to zero to avoid unreliable attention map. As a result, for lost frame it will use the average response of the past non-lost frames. In this block, each frame can only access the past \(N\) frames. The output of the auxiliary branch is denoted as \(X^{A}\in{R}^{C_{A}\times T}\).

The \(X^{S}\) and \(X^{A}\) are merged and fused into \(X^{M}\) by several linear layers before feeding into the vocoder. For synthesis, we employ a causal HiFi-GAN based on the work in \cite{kong2020hifi}. The original HiFi-GAN in \cite{kong2020hifi} operates on mel-spectrogram and targets at 22.05 kHz high-fidelity audio. We modify it to take the latent features \(X^{M}\in{R}^{C_{M}\times T}\) as input and upsample 160 times to match the temporal resolution of a 16kHz signal in a causal way.

\subsection{Contrastive Pretraining}
\label{sssec:contrastive pretraining}
During pre-training, we guide the semantic branch with a contrastive task \(L_{m}\) as shown in Fig.\ref{Fig_scheme_contrastive}. We leverage the target audio as input in the upper branch to get the true quantized latent representation. The packet loss map is taken as the masking scheme. This task requires the lost frames to identify the true quantized latent representation of their targets within a set of distractors from all masked positions. Since the loss patterns of different utterances distribute non-uniformly, we enlarge the candidate list of distractors to the whole batch. Specifically, the distractors are randomly sampled from other lost frames across utterances within the same batch. The contrastive learning is defined as a classification task given by
\begin{equation}\label{(1)}
L_m = -log\frac{exp(sim(x_t,y_t)/\kappa)}{\sum_{\tilde{y}\sim Y^{Q}}exp(sim(x_t, \tilde{y})/\kappa)},
 \end{equation}
 where we compute the cosine similarity between the semantic representations \(x_t\in{X^S}\) and the quantized latent speech representations \(Y^Q\). \(Y^Q\) is a set of \(K+1\) quantized candidate representations including the true target \(y_t\) and \(K\) distractors. \(\kappa\) is the temperature of softmax distribution. 
 
 The quantizer is trained with Gumbel softmax \cite{jang2016categorical} and guided with a diversity loss \(L_d\) as that in \cite{baevski2020wav2vec} to encourage the model to use the codebook entries equally. Specifically, it is defined as a maximization of the entropy of the averaged softmax distribution over the codebook entries for each codebook across a batch, that is
 \begin{equation}\label{(2)}
 L_d = \frac{1}{GV}\sum_{g=1}^{G}\sum_{v=1}^{V}\overline{p}_{g,v}log\overline{p}_{g,v},
 \end{equation}
 where \(G\) and \(V\) are the number of codebooks and entries in each codebook, respectively. \(\overline{p}_{g,v}\) is the averaged possibility to sample the vector \(v\) of group \(g\) over the batch.

 The final contrastive loss is a weighted average of the two terms given by
\begin{equation}\label{(3)}
L_{contrastive}=L_m+\alpha L_d,
 \end{equation}
where $\alpha$ is the wegihting factor.

\subsection{End-to-End Adversarial Training}
\label{sssec:fintuning}
In this stage, we train a generative model conditioned on the semantic features and the auxiliary features to be learned. The semantic branch is freezed without finetuning in this stage. The model is trained in an adversarial manner with two time-domain discriminators: the Multi-Period Discriminator (MPD) and the Multi-Scale Discriminator (MSD) used in HiFi-GAN \cite{kong2020hifi}. For the generator, besides the adversarial loss and feature matching loss, power-law compressed MSE(\(L_{bin}\)) and multi-scale Mel-spectrum MAE (\(L_{mel}\)) from our previous work \cite{xue2022towards} are also used towards better signal fidelity and quality. The final loss is given by
\begin{equation}\label{(4)}
  \begin{split}
        L_{G}=\lambda_{adv}[L_{adv}(G;D) +\lambda_{fm}L_{fm}(G;D)]\\+\lambda_{bin}L_{bin}(G)+\lambda_{mel}L_{mel}(G).  \\
  \end{split}
 \end{equation}
 

\section{Experimental Results}
\label{sec:ExperimentalSetup}
\subsection{Datasets and Settings}
\label{ssec:dataset}
We use the 16kHz raw clean speech from Deep Noise Suppression Challenge at ICASSP 2021 \cite{reddy2021icassp}. It includes multilingual speech and emotional clips. For packet loss, we simulate with a random loss rate from \(10\%\) to \(50\%\), whose maximum burst loss is up to 220ms. Besides, we also simulate a WLAN packet loss pattern with three-state Markov models. For training, we synthesized 600 hours of data with a balanced amount of different loss rate categories. For testing, we use the blind test set from Audio Deep Packet Loss Concealment Challenge at INTERSPEECH 2022 \cite{diener2022interspeech}. This test set includes 966 packet loss traces captured in real Microsoft Teams calls with the maximum burst loss up to 1000ms. To verify the intelligibility, the test-clean dataset from LibriTTS corpus \cite{zen2019libritts} augmented with markov-simulated traces is used for the word error rate (WER) evaluation.

The kernel size of convolutions in feature extractor is \((2,5)\). The stride of the frequency dimension is \((1,4,4,2)\) for 4 convolutional layers. The feature dimensions \(C_{S}\), \(C_{A}\) and \(C_{M}\) are set to 240. The context aggregator is composed of two repeated blocks of TCM modules with dilation rates \((2^0, 2^1, 2^2)\) and a kernel size of 9. The temporal attention window size \(N\) of the auxiliary branch is set to 150. The hyper-parameters of the causal HiFi-GAN network are set as follows: \(h_u=512, k_u=[16,10,4,4], k_r=[3,7,11], D_r=[[1,1],[3,1]]\times 3\). \(h_u, k_u, k_r,D_r\) share the same meaning as that in HiFi-GAN \cite{kong2020hifi}. The total model size is about 33MB, in which the synthesis network accounts for about 26MB. 

For training, we use Adam optimizer with a learning rate of \(2\times10^{-4}\) for contrastive pre-training. Other hyper-parameters in contrastive task are set as \(K=100\), \(\kappa=0.1\), \(G=2\), \(V=320\) and \(\alpha=0.1\). The temperature of Gumbel softmax is annealed from 2 to 0.5 by a factor of 0.999995 at every update \cite{baevski2020wav2vec}. The pretrained model is trained for 240 epochs with a batch size of 256. For the end-to-end adversarial training, the same training configurations as HiFi-GAN \cite{kong2020hifi} are used with weights set as \(\lambda_{adv}=1.0, \lambda_{fm}=2.0, \lambda_{bin}=45, \lambda_{mel}=0.045\). The second stage is trained for 50 epochs with a batch size of 64.
 
\subsection{Evaluation Metrics}
\label{ssec:evaluation}
We evaluate the quality from several perspectives, i.e. signal fidelity, peceptual quality and intelligibility. PESQ and mel cepstral distortion (MCD in dBs) are commonly used to measure the similarity with the target. For perceptual quality, we choose several PLC-related assessment tools. NISQA-Discontinuity \cite{mittag2021nisqa} studied the correlation between the continuity of the transmitted speech and the subjective feelings. PLCMOS \cite{diener2022interspeech} is designed specifically for evaluating PLC algorithms. NISQA-TTS \cite{mittag2021nisqa} is also used to evaluate the naturalness of the synthesized speech as PLC is a kind of synthesis task. Furthermore, to evaluate the intelligibility, we use the pretrained Automatic Speech Recognition (ASR) model on LibriSpeech from WeNet toolkit\cite{zhang2022wenet} to calculate WER.

\subsection{Baselines for Comparison}
\label{ssec:BaselinesForComparison}
To verify the PLC performance, we compare the proposed scheme with the UNet-style structure. Specifically, we implement a frequency-domain generative adversarial network based on TFNet \cite{jiang2022end} with causal convolutional encoder and decoder and skip connections between them. Interleaved structure of TCMs and G-GRUs \cite{jiang2022end} are used to effectively build the context for lost-frames and catch information to reconstruct non-lost frames. The model size is comparable to the proposed scheme with the same input representation. The model is trained to predict the magnitude and phase residual for the lost frames. Loss functions and discriminators are the same as our previous work \cite{xue2022towards}. Furthermore, to verify the effectiveness of contrastive learning, the same structure without contrastive pretraining has also been trained for comparison.

 \begin{figure}[bth]
\centering 
\includegraphics[width=0.5\textwidth]{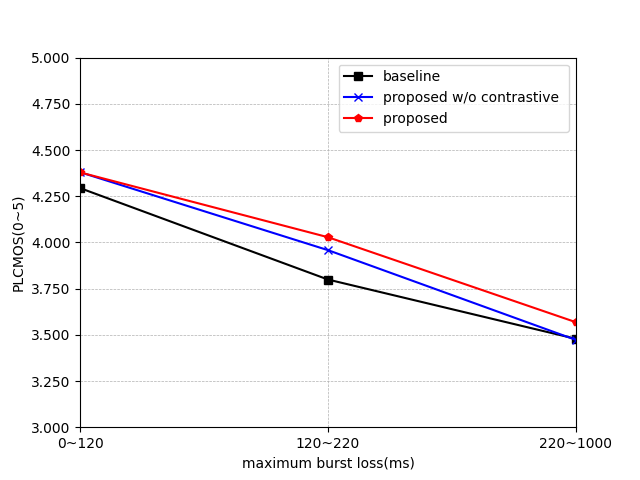} 
\caption{PLCMOS over different traces according to maximum burst loss.}
\vspace{-0.4cm}
\label{Fig2}
\end{figure}

\label{sec:results}
\subsection{Comparison with Other Schemes}
\label{ssec:ComparionWithOtherSchemes}
Tab.\ref{table:tabel1} shows the overall metrics on the blind test set. It can be seen that the proposed scheme ranks the first for all subjective metrics, PLCMOS, NISQA Discontinuity and NISQA TTS, which shows its better concealment ability. Compared with the one without contrastive learning, the proposed scheme generates more natural content with better NISQA Discontinuity and achieves better intelligibility given by lower WER. With regards to PESQ and MCD, it is expected that the proposed synthesis structure is worse than UNet baseline since GAN-based vocoder persues statistical similarity with target speech. For a deep study on the prediction ability, we show its PLCMOS over traces in three categories whose maximum burst loss lies in \([0,120],(120,220],(220,1000] ms\), respectively. As shown in Fig.\ref{Fig2}, all models can do a good job when the loss duration is below 120ms. With longer burst loss, e.g. 140ms, accompanied with more frequent loss, the proposed scheme is able to conceal more plausible content, indicating that it has some sense of semantic prediction ability. Particularly for traces from \((120,220]ms\), where most lost frames lie in a phoneme, the proposed scheme improves a lot against UNet baseline. The results under \((220,1000]ms\) are very close since all models' outputs have passed through the fade-in-fade-out filter after the 140ms-loss. Audio samples could be found at \url{https://contrast-plc.github.io}.

 \begin{table}[h!t]
\vspace{-0.2cm}
\caption{Abaltion study on auxiliary branch}
\begin{tabular}{ c|c|c }
\hline
Scheme&PLCMOS&NISQA TTS\\
\hline
proposed &4.096&3.25\\
w/o auxiliary branch	&3.236&2.500\\
\hline
\end{tabular}
\label{table:tabel2}
\end{table}

\subsection{Ablation Study on Auxiliary Branch}
\label{ssec:AbalationStudy}
To verify the effectiveness of the auxiliary branch, the model with the pretrained semantic branch but without the auxiliary branch is also trained for comparison. We found in Tab.\ref{table:tabel2} that all the metrics drop a lot, indicating that the auxiliary branch is very necessary to make up for the missing low-level information to synthesize the speech with good perceptual quality. From audio samples at \url{https://contrast-plc.github.io}, we can find that most of the content has been concealed but details like timbre and prosody are distorted.

\section{Conclusions}
\label{sec:conclusions}
In this paper, we propose to use contrastive learning to learn loss-robust semantic representation for PLC with a semantic-prediction-and-synthesis network. Lower WER and enhanced prediction ability for longer burst loss(e.g. \(140ms\)) show its effectiveness in phoneme-level understanding which is helpful to the PLC task. The proposed scheme has also surpassed the UNet-style structures where artifacts tend to appear over 120ms.




\bibliographystyle{IEEEbib}
\bibliography{strings,refs}

\end{document}